\documentclass[aps, pre,  showkeys, twocolumn,   showpacs]{revtex4}

\usepackage[small]{caption}
\usepackage{amssymb, epsfig, amsmath,amsfonts, epsf, epsfig}\usepackage{graphicx}

\newcommand{\R}{\mathbb{R}}

\newcommand{\vep}{\varepsilon}

\begin{document}

\title{Modeling Single-File Diffusion by  Step Fractional Brownian Motion and Generalized Fractional Langevin Equation }
\author{S.C. Lim} \address{ Faculty of Engineering,
Multimedia University, Jalan Multimedia, Cyberjaya, 63100, Selangor
Darul Ehsan, Malaysia.}\email{sclim@mmu.edu.my}

\author{L.P. Teo} \address{Faculty of Information
Technology, Multimedia University, Jalan Multimedia, Cyberjaya,
63100, Selangor Darul Ehsan, Malaysia.}\email{lpteo@mmu.edu.my}



\begin{abstract}
Single-file diffusion behaves as normal diffusion at small time and as anomalous subdiffusion at large time. These properties can be described by fractional Brownian motion with variable Hurst exponent or multifractional Brownian motion. We introduce a new stochastic process called Riemann-Liouville step fractional Brownian motion which can be regarded as a special case of multifractional Brownian motion with step function type of Hurst exponent tailored for single-file diffusion. Such a step fractional Brownian motion can be obtained as solution of fractional Langevin equation with zero damping. Various types of fractional Langevin equations and their generalizations are then considered to decide whether their solutions provide the correct description of the long and short time behaviors of single-file diffusion. The cases where dissipative memory kernel is a Dirac delta function, a power-law function, and a combination of both of these functions, are studied in detail.  In addition to the case where the short time behavior of single-file diffusion behaves as normal diffusion, we also consider the possibility of the process that begins as ballistic motion.
\end{abstract}
\maketitle
\section{Introduction}

Single-file diffusion (SFD) refers to the motion of particles in quasi-one-dimensional channels and pores which are so narrow that the particles are unable to pass each other.  The exclusion of a mutual passage of the diffusing particles means that the sequence of particle labels does not change over time. SFD is encountered in many physical, chemical and biological systems, which include the molecular and atomic motion in zeolites   and nanotubes, particle flows in microfluidic devices, ion transport in cell membranes, colloidal motion in narrow tubes, etc. \cite{1,2_1,2_2,2_3,3,4,5}.

The main feature of SFD is that for diffusion time $t$ smaller than typical inter-particle collision time $\tau_c$, the particles diffuse normally and satisfy Fick's law with its mean-square displacement (MSD) $\bar{\Delta}^2(t):=\left\langle \left[x(t) -x(0)\right]^2\right\rangle$ given by
\begin{equation}\label{eq1}
\lim_{t\ll \tau_c}\bar{\Delta}^2(t)=2D_0t,
\end{equation}
with $D_0$  the diffusion coefficient. In other words, for $t\ll \tau_c$, the motion is just ordinary Brownian motion, which is a Markov process. However, for $t\gg \tau_c$,		
\begin{equation}\label{eq2}
\lim_{t\gg \tau_c}\bar{\Delta}^2(t) =2F\sqrt{t},
\end{equation}
where $F$ is the SFD mobility. Recall that diffusion that does not satisfy Fick's law is known as anomalous diffusion with MSD satisfying $\bar{\Delta}^2(t)  \propto t^{\alpha}$, $\alpha\neq 1$. It is called superdiffusion when $\alpha>1$, and subdiffusion when $\alpha<1$. Thus, the long-time behavior of SFD belongs to anomalous subdiffusion, which is non-Markovian, indicating the motion is correlated. Note that SFD displays anomalous  diffusion characteristics even when particle-channel interactions are not taken into account.

Another way to characterize SFD is through its probability density function (or propagator) $P(x,t)$. The probability of finding a particle at position $x$ at time $t$, if it is initially at the origin, approaches the Gaussian propagator after a long time:
\begin{equation*}\label{eq3}
P(x,t) =\frac{1}{\sqrt{2\pi \bar{\Delta}^2(t)}}\exp\left(-\frac{x^2}{2\bar{\Delta}^2(t)}\right).
\end{equation*}
For $t\ll \tau_c$, one has
\begin{equation*}\label{eq4}
P(x,t)=\frac{1}{\sqrt{4\pi D_0t}}\exp\left(-\frac{x^2}{4D_0t}\right),
\end{equation*}
and for $t\gg \tau_c$,
\begin{equation*}\label{eq5_1}
P(x,t)=\frac{1}{\sqrt{4\pi F\sqrt{t}}}\exp\left(-\frac{x^2}{4F\sqrt{t}}\right).
\end{equation*}

	The notion of SFD was first introduced by Hodgkin and Keynes \cite{6} who used it to describe the diffusion of ions through narrow channels in biological membranes. Harris was first to provide a theoretical derivation of \eqref{eq1} and \eqref{eq2} for SFD based on statistical argument \cite{7}. Subsequently, this result was obtained using various models and methods by several authors including Levitt \cite{8}, Fedders \cite{9}, van Beijeren et al \cite{10} and K\"arger \cite{11}. Recently, there are also attempts to model SFD based on fractional diffusion equations and fractional Langevin equations \cite{a,b,c,d}.  Despite of the numerous theoretical models and numerical simulations, experimental evidence for the occurrence of SFD was only obtained quite recently \cite{12,13,14,15,16}. The main reason is that there is a lack of ideal experimentally accessible single file systems.	

	The main aim of this paper is to propose  some  stochastic processes to describe the SFD. We do not address the detailed mechanism of SFD, but instead we emphasize more on the possibility of finding    random processes which have the basic properties of SFD.
Various types of fractional Langevin equations are considered in order to see whether they yield the stochastic processes which satisfy the basic statistical properties of SFD.

\section{Modeling Single-File Diffusion by Step Fractional Brownian Motion}
In this section we introduce a generalization of standard fractional Brownian motion (FBM) \cite{17} called step fractional Brownian motion (SFBM) and show that it can be used to describe the basic statistical properties of SFD. FBM has been widely used to model many areas such as turbulence, internet traffic, financial time series, biomedical processes, etc. One limitation of FBM model is that the long time (or low frequency) behavior that exhibit long-range dependence, and the short time (or high frequency) behavior that characterizes fractal property are both described by a single Hurst parameter $H$. Furthermore, a constant Hurst parameter is too restrictive for many applications. During the past decade different generalizations of FBM have been proposed to address this problem. Among them the most well-known is the multifractional Brownian motion (MBM), which was introduced independently in \cite{18} and \cite{19}. For MBM the Hurst parameter $H$ is replaced by $H(t)$, a deterministic function depending   on time. MBM was later extended to generalized multifractional Brownian motion (GMBM) in order to model systems which require $H(t)$ to be a irregular function of time \cite{20_1, 20_2}. However, there exist processes and phenomena which exhibit abrupt changes of Hurst parameter requiring $H(t)$ to be a piecewise constant function of time. For the description of such behavior, Benassi et al \cite{21,22} introduced the step fractional Brownian motion (SFBM). A similar process known as multiscale fractional Brownian motion with its Hurst parameter varying as a piecewise function of frequencies was also studied by several authors \cite{23,24}.

Recall that the standard FBM  $B_H(t)$ is a Gaussian process with mean zero and correlation function given by
\begin{equation*}\label{eq5_2}
\left\langle B_H(t) B_H(s)\right\rangle =\frac{C_H}{2}\left(|t|^{2H}+|s|^{2H}-|t-s|^{2H}\right),
\end{equation*}
where $$C_H=\frac{\Gamma(1-2H)\cos(\pi H) }{\pi H}.$$ $B_{H}(t)$   is not a stationary process, but its increment process is stationary. FBM  is a self-similar process which satisfies for all $a\in \mathbb{R}_+$,
\begin{equation*}\label{eq6}
B_H(at) \triangleq a^HB_H(t),
\end{equation*}
where $\triangleq$  denotes equality in all finite distributions. The stationary property of the increments of $B_H(t)$    allows the following harmonizable representation for the process:
\begin{equation*}\label{eq7}
B_H(t)=\frac{1}{\sqrt{2\pi}}\int_{-\infty}^{\infty} \frac{e^{i\omega t}-1}{|\omega|^{H+1/2}}\tilde{\eta}(\omega)d\omega,
\end{equation*}
where $0<H<1$, $t\in\R$   and  $\tilde{\eta}(\omega)$ is the Fourier transform of $\eta(t)$ --- the standard white noise defined by
\begin{equation*}\label{eq8}
\langle \eta(t)\rangle =0, \hspace{1cm} \left\langle \eta(t)\eta(s)\right\rangle =\delta(t-s).
\end{equation*}

For modeling a process that evolves from time $t=0$, instead of using the usual or standard FBM (which begins at time $t=-\infty$), it will be more appropriate to use an alternative FBM that starts at time zero. This second type of FBM is known as Riemann-Liouville FBM (RL-FBM), which is defined as the RL fractional integral of white noise \cite{25}:
\begin{equation}\label{eq9}
W_H(t) = \frac{1}{\Gamma\left(H+\frac{1}{2}\right)}\int_0^t (t-u)^{H-\frac{1}{2}}\eta(u)du, \;\; t\in \R_+, \;H>0.
\end{equation}
 $W_H(t)$   is a Gaussian process with zero mean  $\langle W_H(t)\rangle=0$ and correlation function given by
 \begin{equation*}\label{eq10}\begin{split}
 &C_{W_H}(t,s) =\left\langle W_H(t)W_H(s)\right\rangle \\=& \frac{t^{H-1/2}s^{H+1/2}}{\left(H+\frac{1}{2}\right)\Gamma\left(H+\frac{1}{2}\right)^2}\,_2F_1\left(\frac{1}{2}-H,1,H+\frac{3}{2}; \frac{s}{t}\right)
 \end{split}\end{equation*}
when $s<t$.   Here $_2F_1(a,b,c;z)$  denotes the Gauss hypergeometric function. The variance  of the process $W_H(t)$ is
\begin{equation}\label{eq11}
Var(W_H(t))=\left\langle W_H(t)^2\right\rangle =\frac{t^{2H}}{2H \Gamma\left(H+\frac{1}{2}\right)^2}.
\end{equation}

Note that for the standard FBM $B_H(t)$, the  Hurst parameter $H$ should lie in the range $(0,1)$, whereas for RL-FBM $W_H(t)$, $H$  takes any positive real value. Both $B_H(t)$ and $W_H(t)$   reduce to ordinary Brownian motion when $H=1/2$. In contrast to $B_H(t)$  which has stationary increments, the increments of  $W_H(t)$ are non-stationary. Due to the failure of its increments to be stationary, $W_H(t)$  does not has a harmonizable representation. When $t\rightarrow \infty$, RL-FBM approaches the standard FBM \cite{25}.

	Now we want to consider the step fractional Brownian motion (SFBM). Such a generalization of FBM was first introduced for standard FBM using the harmonizable representation as follow \cite{21}:
\begin{equation*}\label{eq12}
B_{H(t)}(t)=\frac{1}{\sqrt{2\pi}} \int_{-\infty}^{\infty}\frac{e^{i\omega t}-1}{|\omega|^{H(t)+1/2}}\tilde{\eta}(\omega)d\omega,
\end{equation*}
where
\begin{equation}\label{eq13}H(t)=\sum_{i=1}^N \boldsymbol{1}_{[\tau_{i-1}, \tau_{i})}H_i,                       				  	    \end{equation}
 with $\tau_0=-\infty$  and $\tau_{N}=\infty$, $H_i\in (0,1)$, $\boldsymbol{1}_I(t)=1$ if $t\in I$ and $\boldsymbol{1}_I(t)=0$ if $t\notin I$. This is an adaptation of MBM, which is defined for a time-dependent Hurst parameter $H(t)$. Due to the absence of such representation in RL-FBM, we generalize $W_H(t)$   to RL-SFBM based on the moving average representation \eqref{eq9}:
\begin{equation}\label{eq14}
W_{H(t)}(t) =\frac{1}{\Gamma\left(H(t)+\frac{1}{2}\right)}\int_0^t (t-u)^{H(t)-1/2}\eta(u)du,
\end{equation}
with $H(t)$   the piecewise function given by \eqref{eq13}, except that in this case $\tau_0=0$, and $H_i\in (0, \infty)$. Its covariance  is given by
\begin{equation}\label{eq15}\begin{split}
&C_{W_H(\cdot)}(t,s) =\left\langle W_{H(\cdot)}(t)W_{H(\cdot)}(s)\right\rangle \\=& \frac{t^{H(t)-1/2}s^{H(s)+1/2}}{ \Gamma\left(H(s)+\frac{3}{2}\right)\Gamma\left(H(t)+\frac{1}{2}\right)}\\&\times\,_2F_1\left(\frac{1}{2}-H(t),1,H(s)+\frac{3}{2}; \frac{s}{t}\right),
\end{split}
\end{equation}if $s<t$, and its variance $\left\langle W_{H(\cdot)}(t)^2\right\rangle $ is given by \eqref{eq11} with $H$ replaced by $H(t)$.

The property of global self-similarity does not apply to both MBM and SFBM. In the case of MBM, the notion of self-similarity is replaced by the local asymptotic self-similarity \cite{19}, which is also satisfied by SFBM and RL-SFBM with some modification. Suppose $W_{H(t)}(t)$   is a RL-SFBM  with scaling function  $H(t)$ defined above. For all $t\in (\tau_{i-1}, \tau_{i})$,
\begin{equation*}
\label{eq16}\begin{split}
&\lim_{\vep\rightarrow 0}\left\{\frac{W_{H(\cdot)}(t+\vep u)-W_{H(\cdot)}(t)}{\vep^{H_i}}\right\}_{u\in \R_+}\\\triangleq &\left\{B_{H_i}(u)\right\}_{u\in \R_+}
\end{split}
\end{equation*}
The convergence is in the sense of distributions. In other words, the tangent process of RL-SFBM for each scale $H_i$  is $B_{H_i}$, a FBM indexed by $H_i$.

\begin{figure}\centering \epsfxsize=0.8\linewidth
 \epsffile{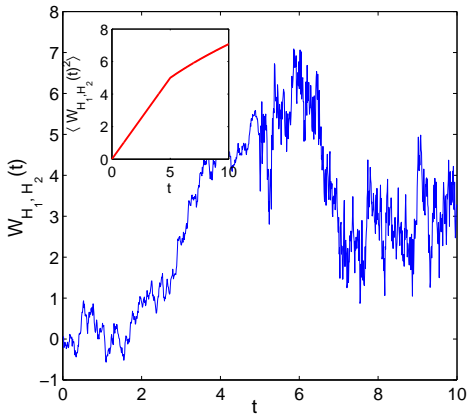} \caption{\label{f1}  Two step RL-SFBM $W_{H_1,H_2}(t)$ with $H_1=0.5$, $H_2=0.25$, $\tau=5$, $\chi_{H_1}=1$ and $\chi_{H_2}^2= \sqrt{5}\Gamma(0.75)^2/2$. The smaller window shows the MSD of the process. }\end{figure}

For modeling SFD, we use RL-SFBM with single change of scale, that is the process \eqref{eq14} with $H(t)=H_1\boldsymbol{1}_{[0,\tau)}+H_2\boldsymbol{1}_{[\tau,\infty)}$. To be more specific, we denote this process by $W_{H_1, H_2}(t)$, the two-scale RL-SFBM indexed by $H_1>0$ and $H_2>0$. For this simple case, we can write
\begin{equation}\label{eq17}\begin{split}
W_{H_1, H_2}(t) =& \frac{\boldsymbol{1}_{[0, \tau)}(t)}{\Gamma\left(H_1+\frac{1}{2}\right)}\int_0^t (t-u)^{H_1-\frac{1}{2}}\eta(u)du \\
&+\frac{\boldsymbol{1}_{[\tau, \infty)}(t)}{\Gamma\left(H_2+\frac{1}{2}\right)}\int_{0}^t (t-u)^{H_2-\frac{1}{2}}\eta(u)du.
\end{split}
\end{equation}
 From the above definition, one sees that  $W_{H_1, H_2}(t)$ is a Gaussian process with zero mean and correlation function
\begin{equation*}\label{eq18}\begin{split}
&\left\langle W_{H_1,H_2}(t)W_{H_1,H_2}(s)\right\rangle \\=&\boldsymbol{1}_{[0,\tau)}(t)\boldsymbol{1}_{[0,\tau)}(s)C_{H_1, H_1}(t,s)\\&+\boldsymbol{1}_{[\tau,\infty)}(t)
\boldsymbol{1}_{[0,\tau)}(s)C_{H_1,H_2}(t,s)\\&+\boldsymbol{1}_{[\tau,\infty)}(t)\boldsymbol{1}_{[\tau,\infty)}(s) C_{H_2,H_2}(t,s)
\end{split}\end{equation*}
when $s<t$, and $C_{H_i,H_j}(t,s)$, $i,j=1,2$, is given by \eqref{eq15} with $H(t)=H_i$ and $H(s)=H_j$. Similarly, its variance $\left\langle W_{H_1,H_2}(t)^2\right\rangle $ is given by \eqref{eq11} with $H$ replaced by $H(t)$.
Thus, $W_{H_1, H_2}(t)$, the RL-SFBM with two scales, behaves like $W_{H_1}(t)$  for $t\in [0,\tau)$, and behaves like $W_{H_2}(t)$    when $t\in [\tau,\infty)$   (see FIG. \ref{f1}). $W_{H_1H_2}(t)$  is piecewise self-similar, it is self-similar of order $H_1$  in the time interval $[0,\tau)$, and of order  $H_2$ in the time interval $[\tau,\infty)$.

In order to use RL-SFBM for modeling SFD, it is necessary to carry out some minor modifications to the definition of $W_{H_1,H_2}(t)$.  \eqref{eq14} and \eqref{eq17} are to be replaced by
\begin{equation}\label{eq20}
W_{H(t)}(t) =\frac{\chi_{H(t)}}{\Gamma\left(H(t)+\frac{1}{2}\right)}\int_0^t (t-u)^{H(t)-1/2}\eta(u)du,
\end{equation}
and
\begin{equation}\label{eq21}\begin{split}
W_{H_1, H_2}(t) =& \frac{\chi_{H_1}\boldsymbol{1}_{[0, \tau)}(t)}{\Gamma\left(H_1+\frac{1}{2}\right)}\int_0^t (t-u)^{H_1-\frac{1}{2}}\eta(u)du \\
&+\frac{\chi_{H_2}\boldsymbol{1}_{[\tau, \infty)}(t)}{\Gamma\left(H_2+\frac{1}{2}\right)}\int_{0}^t (t-u)^{H_2-\frac{1}{2}}\eta(u)du,
\end{split}
\end{equation}
where $\chi_{H_i}$, $i=1,2$, are positive constants which are introduced for the purpose of obtaining the correct coefficients for the MSD of the diffusing particles. Note that by definition, $\langle W_{H(\cdot)}(t)\rangle =0$ and $W_{H(\cdot)}(0)=0$. Therefore the variance of $W_{H(\cdot)}(t)$ is equal to the MSD.

\begin{figure}\centering \epsfxsize=0.8\linewidth
 \epsffile{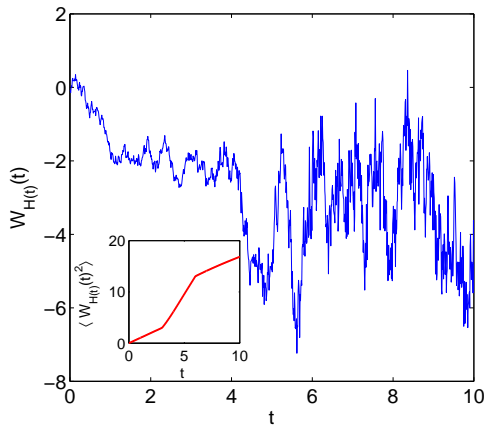} \caption{\label{f4}  Piecewise linear RL-MBM $W_{H(t)}(t)$ with $H(t)=0.5$, for $t\in [0,3]$, $H(t)=0.25$ for $t\in [6,10]$ and $H(t)$ is linear in for $t\in [3,6]$.   $\chi_{H(t)}=1$ for $t\in [0,3]$ and $\chi_{H(t)}= 2$ for $t\in [6,10]$. For $t\in [3,6]$, $\chi_{H(t)}$ is linear. The smaller window shows the MSD of the process. }\end{figure}

Now we want to see how the RL-SFBM with two scales can be used to described the basic properties of SFD. By letting $H_1=1/2$, $H_2=1/4$, and $\chi_{H_i}$   in terms of diffusion coefficient  $D_0$ and SFD mobility $F$ with $\chi_{H_1}=2D_0$, $\chi_{H_2}=\Gamma(0.75)^2F$, then the MSD of $W_{H_1, H_2}(t)$ \eqref{eq21} is equal to  \begin{equation*}\label{eq19}\begin{split}
\left\langle W_{H_1,H_2}(t)^2\right\rangle=&\left\langle\left[W_{H_1H_2}(t)-W_{H_1H_2}(0)\right]^2\right\rangle \\=&\begin{cases}
2D_0 t, \hspace{1cm}& t\in [0,\tau)\\
2F\sqrt{t}, &t\in[\tau,\infty),
\end{cases}
\end{split}\end{equation*}which is the same as the MSD of SFD.  The value of $\tau$ is set to equal to $\sqrt{F/D_0}$ so that the MSD is continuous. A sample path of the process $W_{H_1, H_2}(t)$ is shown in FIG. \ref{f1}.

There are some further generalizations of RL-SFBM which can be used to model SFD. For example, one can use a piecewise linear function $H(t)$ in \eqref{eq20} so that $H(t)=1/2$ for $t\in [0, \tau_1)$, $H(t)=1/4$ for $t\in [\tau_2,\infty)$ and $H(t)$ is a linear function interpolating the points $(\tau_1, 1/2)$ and $(\tau_2, 1/4)$ in the interval $[\tau_1, \tau_2]$. If $\chi_{H(t)}=2D_0$ for $t\in [0, \tau_1)$, $\chi_{H(t)}=\Gamma(0.75)^2F$ for $t\in [\tau_2,\infty)$,   and $\chi_{H(t)}$ is a linear function of $t$ in the interval $[\tau_1,\tau_2]$ so that $\chi_{H(t)}$ is continuous, then the   process $W_{H(t)}(t)$  is a special case of RL-MBM, which provides a model for SFD that has continuous sample paths (see FIG. \ref{f4}).

\begin{figure}\centering \epsfxsize=0.8\linewidth
 \epsffile{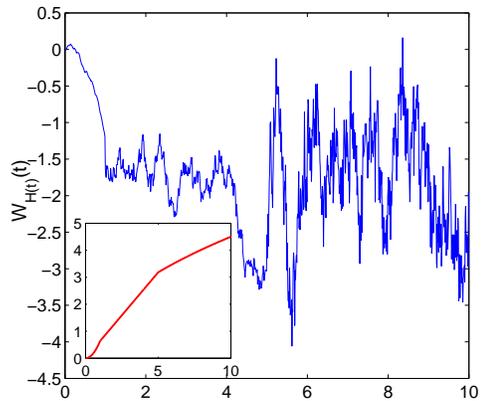} \caption{\label{f7}  Three step RL-SFBM $W_{H_1,H_2, H_3}(t)$ with $H_1=1$, $H_2=0.5$, $H_3=0.25$, $\tau_1=1$, $\tau_2=5$, $\chi_{H_1}=1$, $\chi_{H_2}^2=1/(2\Gamma(1.5)^2)$ and $\chi_{H_3}^2= \sqrt{5}\Gamma(0.75)^2/(4\Gamma(1.5)^2)$. The smaller window shows the MSD of the process. }\end{figure}

	As remarked by K\"arger \cite{2_1},   random walk model can only be regarded as an approximation to the real SFD systems. At short times, such systems first undergo ballistic motion with the MSD $\bar{\Delta}(t)\sim t^2$. In other words, there is the possibility of the direct
transition from the ballistic regime to the single-file regime. Such a tendency becomes more prominent with increasing concentration; and it has been demonstrated by molecular dynamical simulations \cite{x}. For modeling SFD that is ballistic at small $t$, we can use RL-SFBM with three scales.  Here we consider the case of a SFD process with three regimes: initial ballistic regime followed by the normal diffusion, and finally the single-file diffusion region. For such a process, we let the time-dependent Hurst exponent in \eqref{eq20} to be $$H(t) =H_1 \boldsymbol{1}_{[0,\tau_1)}(t) + H_2 \boldsymbol{1}_{[ \tau_1,\tau_2)}(t)+H_3\boldsymbol{1}_{[ \tau_2, \infty)}(t),$$ where $H_1=1$, $H_2=1/2$ and $H_3=1/4$. An example of such a process is shown in FIG. \ref{f7}.

\section{Modeling Single-File Diffusion by Fractional Langevin Equations}
In this section we want to examine whether it is possible to describe the basic characteristics of SFD based on the various types of fractional Langevin equations. We shall first show that how RL-SFBM can be obtained as the solution to the "free fractional Langevin equation" (fractional Langevin equation without damping). This will be followed by discussion on the  fractional generalized Langevin equation and its extensions.

\subsection{ Fractional Langevin equation}\label{s3a}
First note that the definition of RL-FBM \eqref{eq9} can be written as:
\begin{equation}\label{eq22}
W_H(t) =\,_0I_t^{H+1/2}\eta(t),
\end{equation}
where the Riemann-Liouville (RL) fractional integral $_0I_t^{\alpha}$  is given by \cite{26,27,28,29}:
\begin{equation*}\label{eq23}\begin{split}
(\,_0I_t^{\alpha}f)(t) =\frac{1}{\Gamma(\alpha)}\int_0^t(t-u)^{\alpha-1}f(u)du.
\end{split}
\end{equation*}The Riemann-Liouville fractional derivative $_0D_t^{\alpha}$ is defined as
\begin{equation}\label{eq4_23_1}_0D_t^{\alpha}:=D_t^n \, _0I_t^{n-\alpha}\end{equation}
for $n-1\leq \alpha<n$.
In view of the property that
\begin{equation*}
\begin{split}
&_0I_t^{\alpha}\,_0D_t^{\alpha}f(t)=\,_0I_t^{\alpha}D_t^n \, _0I_t^{n-\alpha}f(t)\\=&f(t)-\sum_{k=1}^n \frac{t^{\alpha-k}}{\Gamma(\alpha-k+1)}\left[\,_0D_t^{\alpha-k}f\right](0),
\end{split}
\end{equation*}
we can consider \eqref{eq22} as  the solution of the "free" fractional Langevin equation
\begin{equation*}\label{eq24}
\begin{split}
_0D_t^{H+1/2}W_H(t)=\eta(t),
\end{split}
\end{equation*}
subject to the initial condition  $(\,_0D_t^{H-1/2}W_H)(0)=0$. In a similar way, we see that \eqref{eq17} can be re-expressed as
\begin{equation*}\label{eq25}\begin{split}
&W_{H_1,H_2}(t)=\,_0I_t^{H(t)+1/2}\eta(t)\\=&\left[\boldsymbol{1}_{[0,\tau)}(t) \,_0I_t^{H_1+1/2}+\boldsymbol{1}_{[\tau,\infty)}(t)\,_0I_t^{H_2+1/2}\right]\eta(t),
\end{split}\end{equation*}
and $W_{H_1,H_2}(t)$  can be regarded as the solution of
\begin{equation*}\label{eq26}
_0D_t^{H(t)+1/2}W_{H(t)}(t)=\eta(t),
\end{equation*}
with $H(t)=H_1\boldsymbol{1}_{[0,\tau)}+H_2\boldsymbol{1}_{[\tau,\infty)}$, and subject to the initial condition   $\left(\,_0D_t^{H(t)-1/2}W_{H(t)}\right)(0)=0$.

Recall that Brownian motion can also be regarded as the position process associated with the solution (velocity process) of the usual Langevin equation. It is natural to ask if a similar link exist for RL-FBM  and RL-SFBM with fractional Langevin equation. For this purpose we first consider the following general type of fractional Langevin equation with two different fractional orders $\alpha$  and $\gamma$  \cite{30}:	
\begin{equation}\label{eq27}
\begin{split}\left(\,_0D_t^{\alpha}+\lambda\right)^{\gamma}v_{\alpha,\gamma}(t) =\eta(t), \;\; 0<\alpha<1, \gamma>0,
\end{split}
\end{equation}
where $\lambda>0$  is the dissipative    parameter, and $\eta(t)$ is the standard white noise. Here we remark that $\left(_0D_t^{\alpha}+\lambda\right)^{\gamma}$   can be regarded as "shifted" fractional derivative as compared with the un-shifted one $_0D_t^{\alpha\gamma}$. By using binomial expansion, the shifted fractional derivative can be formally expressed as
\begin{equation*}\label{eq29}
\left(_0D_t^{\alpha}+\lambda\right)^{\gamma}=\sum_{j=0}^{\infty}\begin{pmatrix} \gamma\\j\end{pmatrix}\lambda^j \;_0D_t^{\alpha(\gamma-j)}.
\end{equation*}
Special case of \eqref{eq27} with $\alpha>0, \gamma=1$   has been considered previously in \cite{32,33,34,4_24_1}; and the solutions of the fractional Langevin equation with $\alpha=1, \gamma>0$    have also been studied in \cite{35, 36}. It is found that for the general case, if one considers the solution $v_{\alpha,\gamma}(t)$ to \eqref{eq27} as the velocity process, then the corresponding position process $x_{\alpha,\gamma}(t)$  depends only on the differential relation between $v_{\alpha,\gamma}(t)$ and  $x_{\alpha,\gamma}(t)$ \cite{31}. If  the usual velocity-position relation is used, that is velocity is the ordinary derivative of position, then the variance of the position process does not depend on the long-time behavior of the correlation of the velocity process. One always get the variance of $x_{\alpha,\gamma}(t)$ behaves like $Var(x_{\alpha,\gamma}(t))\sim t$,  just like the case of normal diffusion. The long-time dependence of the covariance of $v_{\alpha,\gamma}(t)$    which varies as $t^{-\alpha-1}$  does not enter in the leading term of the variance of $x_{\alpha,\gamma}(t)$, it only appears as second leading term \cite{31}.

On the other hand, if one assumes
\begin{equation*}\label{eq30}v_{\alpha,\gamma}(t)=\,_0D_t^{\beta}x_{\alpha,\gamma;\beta}(t), \hspace{0.5cm}\left(_0D_t^{\beta-1}x_{\alpha,\gamma;\beta}\right)(0)=0,
\end{equation*}
where $0<\beta\leq 1$, such that
\begin{equation*}\label{eq31}
x_{\alpha,\gamma;\beta}(t) =\, _0I_t^{\beta}v_{\alpha,\gamma}(t),
\end{equation*}
  then we have shown in \cite{31} that $Var(x_{\alpha,\gamma;\beta}(t))\sim t^{2\beta-1}$.   Therefore for $\beta=3/4$, one gets the correct long time behavior for the MSD of SFD. Here we have used the fact that  $x_{\alpha,\gamma;\beta}(0)$=0. As for the short time behavior, it can be shown   that $Var(x_{\alpha,\gamma;\beta}(t))\sim t^{2\alpha\gamma+2\beta-1}$ \cite{31}. Therefore, in order to obtain the correct short time behavior for the MSD of SFD, we require $\alpha\gamma=1-\beta=1/4$. We notice that for the characterization of the mean square displacements of SFD, the parameters $\alpha$ and $\gamma$ appears in the combination $\alpha\gamma$. Therefore, we can restrict ourselves to the case $\gamma=1$, for then the process $x_{\alpha,1;\beta}(t)$ satisfies the following fractional Langevin equation:
\begin{equation}\label{eq4_16_1}\begin{split}
_0D_t^{\beta}x_{\alpha,1;\beta}(t)=&v_{\alpha,1}(t)\\
\,_0D_t^{\alpha}  v_{\alpha,1}(t)+\lambda v_{\alpha,1}&(t) =\eta(t),
\end{split}
\end{equation}with initial conditions $(_0D_t^{\beta}x_{\alpha;1;\beta})(0)=0$ and $(_0D_t^{\alpha}v_{\alpha;1})(0)=0$.  Setting $\alpha=1/4$ and $\beta=3/4$ recovers the basic properties for SFD. On the other hand, setting $\alpha=\beta=3/4$ gives a process $x_{\alpha,1;\beta}$ that is ballistic (i.e., its MSD is $\sim t^2$) at small $t$, and sub-diffusive with exponent $1/2$ when $t$ is large.
\begin{figure}\centering \epsfxsize=0.8\linewidth
 \epsffile{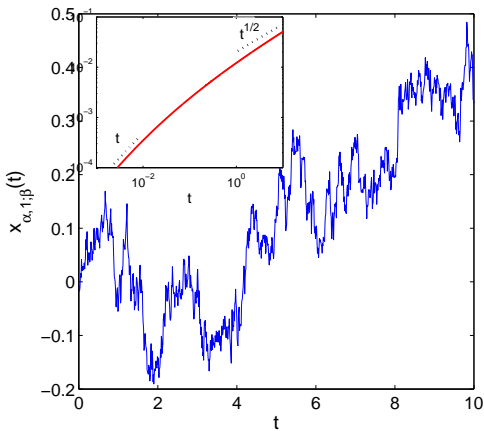} \caption{\label{f5}  A sample path of $x_{\alpha,1;\beta}(t)$ where $\alpha =1/4$ and $\beta=3/4$. The smaller window shows the MSD of the process. }\end{figure}

\begin{figure}\centering \epsfxsize=0.8\linewidth
 \epsffile{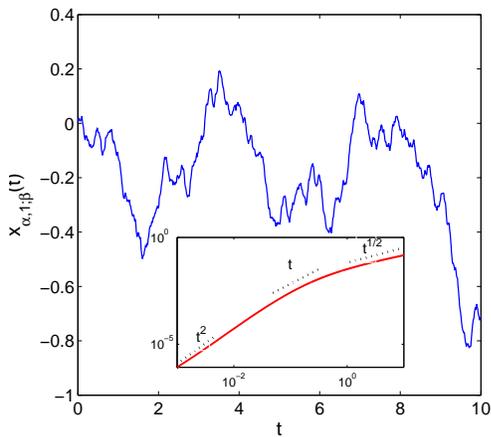} \caption{\label{f6}  A sample path of $x_{\alpha,1;\beta}(t)$ where $\alpha =3/4$ and $\beta=3/4$. The smaller window shows the MSD of the process. }\end{figure}

 Notice that for $n-1\leq \alpha<n$, the Laplace transform of the Riemann-Liouville fractional derivative $_0D_t^{\alpha}$ is given by \cite{27}:
\begin{equation}\label{eq4_23_2}
\widetilde{_0D_t^{\alpha}f}( s)=s^{\alpha}\tilde{f}(s)-\sum_{k=0}^{n-1} s^{k}\left[\,_0D_t^{\alpha-k-1}f\right](0),
\end{equation}where  $\tilde{f}(s)$  denotes Laplace transform of  $f(t)$.  In solving the fractional Langevin equation \eqref{eq27}, we have assumed that $(_0D_t^{\alpha-1}v_{\alpha,\gamma})(0)=0$ so that the Laplace transform of the solution $\tilde{v}_{\alpha,\gamma}(s)$ satisfies
$$\tilde{v}_{\alpha,\gamma}(s)=\frac{\tilde{\eta}(s)}{(s^{\alpha}+\lambda)^{\gamma}},$$ and the solution $v_{\alpha,\gamma}(t)$ is obtained by taking the inverse Laplace transform of this equation. From the practical point of view, the applicability of Riemann-Liouville fractional derivative is limited by the absence of physical interpretation of the initial condition of the type $(_0D_t^{\alpha-1}v_{\alpha,\gamma})(0)=v_0$, when $\alpha$ is not an integer. There is another definition of fractional derivative called Caputo fractional derivative which is defined as
$$^C_0D_t^{\alpha}f(t):=\, _0I_t^{n-\alpha} D_t^n$$ when $n-1<\alpha\leq n$. The difference between the Caputo fractional derivative and the Riemann-Liouville fractional derivative \eqref{eq4_23_1} lies in the order of taking differentiation and integration. In contrast to \eqref{eq4_23_2}, the Laplace transform  of  Caputo fractional derivative is given by  \cite{27}
\begin{equation}\label{eq4_23_3}
\widetilde{^C_0D_t^{\alpha}f}( s)=s^{\alpha}\tilde{f}(s)-\sum_{k=0}^{n-1} s^{\alpha-k-1}f^{(k)}(0),
\end{equation}when $n-1<\alpha \leq n$. The initial conditions that need to be specified now are the values of the ordinary derivatives of $f$ at $t=0$, which have natural physical interpretations. For the system \eqref{eq4_16_1}, we have assumed that  $(_0D_t^{\beta}x_{\alpha;1;\beta})(0)=0$ and $(_0D_t^{\alpha}v_{\alpha;1})(0)=0$, which give the relation
\begin{equation*}\begin{split}
s^{\beta}\tilde{x}_{\alpha,1;\beta}(s) = \tilde{v}_{\alpha,1}(s),\\
s^{\alpha}\tilde{v}_{\alpha,1}(s)+\lambda \tilde{v}_{\alpha,1}(s)=\tilde{\eta}(s)
\end{split}
\end{equation*}for the Laplace transforms of $x_{\alpha,1;\beta}(t)$ and $v_{\alpha,1}(t)$. Comparing the Laplace transforms of the Riemann-Liouville fractional derivative and Caputo fractional derivative \eqref{eq4_23_2} and \eqref{eq4_23_3}, we find that we can also interpret our solution to \eqref{eq4_16_1} as satisfying
\begin{equation}\label{eq5_1_1}\begin{split}
^C_0D_t^{\beta}x_{\alpha,1;\beta}(t)=&v_{\alpha,1}(t)\\
\,^C_0D_t^{\alpha}  v_{\alpha,1}(t)+\lambda v_{\alpha,1}&(t) =\eta(t),
\end{split}\end{equation}with initial conditions $x_{\alpha,1;\beta}(0)=0$ and $v_{\alpha,1}(0)=0$, where now Caputo fractional derivative is used. In the following, we are only going to use Caputo fractional derivative. Therefore, we are going to  use the symbol $_0D_t^{\alpha}$ instead of the symbol $_0^CD_t^{\alpha}$ for Caputo fractional derivative and it should not incur any confusion.
The short time behavior of MSD is sensitive to the initial conditions. For example, the stochastic process satisfying  \eqref{eq5_1_1} with $\alpha=\beta=1$ and initial conditions $v(0)\neq 0$ has MSD behaves like $\sim t^2$ as $t\rightarrow 0$. Whereas for $v(0)=0$, the MSD behaves like $\sim t^3$ as $t\rightarrow 0$.
We will   consider the more general case where $x(0)\neq 0$ and $v(0)\neq 0$ in the subsequent discussions.
\subsection{   Fractional Generalized Langevin Equation}\label{s3b}
We consider the generalized Langevin equation:
\begin{equation}\label{eq32}
\begin{split}&D_tx(t)=v(t),\\
& D_t v(t) +\int_0^t \gamma(t-u)v(u)du=F(t),\\
\end{split}
\end{equation}
where  $\gamma(t)$ is the dissipative   memory kernel, and $F(t)$  is a Gaussian random force with zero mean and correlation
\begin{equation}\label{eq33}
\left\langle F(t)F(s)\right\rangle = C_F(|t-s|).
\end{equation}
When $\gamma(t)=\lambda\delta(t)$ and $C_F(|t|)=\delta(t)$, \eqref{eq32} reduces to the ordinary Langevin equation. It has been shown that it is possible to obtain the position process $x(t)$  as an anomalous diffusion if $F(t)$  is considered as internal noise with long-tailed correlation \cite{37}. More precisely, if  $\gamma(t)=\lambda t^{-\kappa}$, where $0<\kappa\leq 1$,  and fluctuation-dissipation theorem holds with $C_F(t)=k_BT\gamma(t)$, one then has $\langle [x(t)-x(0)]^2\rangle \sim t^{\kappa}$ as $t\rightarrow \infty$. On the other hand, if $F(t)$  is an external noise with $C_F(t)=c_{\theta}|t|^{-\theta}$, $1<\theta<1$, one has $\left\langle x(t)^2\right\rangle \sim t^{2\kappa-\theta}$  as $t\rightarrow \infty$ if $2\kappa>\theta$ \cite{4_16_1, 4_17_2}. We thus have subdiffusion when  $0<2\kappa-\theta<1$, superdiffusion when $2\kappa-\theta>1$, and normal diffusion for $2\kappa-\theta=1$. However, the short time behavior of $x(t)$  is always ballistic \cite{37}.

	In this subsection we want to examine whether it is possible to describe the basic characteristics of SFD based on the fractional Langevin equation in the following general setting:
\begin{equation}\label{eqs1}
\begin{split}
&_0D_t^{\beta}x(t)=v(t), \hspace{1cm} 0<\beta\leq 1,\\
&_0D_t^{\alpha}v(t) +\int_0^t \gamma(t-u) v(u)du=F(t),\hspace{0.5cm} 0<\alpha\leq 1,
\end{split}
\end{equation}
where   now $_0D_t^{\alpha}$ and $_0D_t^{\beta}$ are Caputo   fractional derivatives,  $\gamma(t)$  is the frictional kernel and $F(t)$  is a Gaussian noise with zero mean and the following correlation:
\begin{equation*}\label{eqs3}
\langle F(t)F(s)\rangle = C_F(t-s)=c_{\theta}|t-s|^{-\theta}.
\end{equation*}
When $\beta=1$, we have ordinary velocity, and for $\beta\neq 1$, the velocity is a fractional velocity.

	If the dissipative memory  kernel is given by $\gamma(t)=\lambda t^{-\kappa}$, $0<\kappa\leq 1$, the second equation in \eqref{eqs1} can be written in a more compact form:
\begin{equation}\label{eqs4}
_0D_t^{\alpha}v(t)+\chi \,_0I_t^{\zeta}v(t)=F(t)
\end{equation}
where $\zeta=1-\kappa$ and $\chi=\Gamma(\zeta) \lambda$.  Formally, when  $\zeta=0$ and $F(t)=\eta(t)$, \eqref{eqs4} reduces to ordinary Langevin equation. Laplace transforms of \eqref{eqs1} and \eqref{eqs4} give
\begin{equation}\label{eqs5}
\begin{split}
&s^{\beta}\tilde{x}(s)-s^{\beta-1}\tilde{x}_0=\tilde{v}(s),\\
&s^{\alpha}\tilde{v}(s)-s^{\alpha-1}\tilde{v}_0+\chi s^{-\zeta}\tilde{v}(s) =\tilde{F}(s),
\end{split}
\end{equation}
where    $x_0=x(0)$ and $v_0=v(0)$. We assume that $v_0\neq 0$.   From \eqref{eqs5} one gets
\begin{equation}
\label{eqs7}\begin{split}
\tilde{v}(s) =& \frac{\tilde{F}(s)}{s^{\alpha}\left(1+\chi s^{-\zeta-\alpha}\right)}+\frac{v_0}{s\left(1+\chi s^{-\zeta-\alpha}\right)},\\
\tilde{x}(s) =&\frac{x_0}{s}+\frac{\tilde{F}(s)}{s^{\alpha+\beta}\left(1+\chi s^{-\zeta-\alpha}\right)}+\frac{v_0}{s^{\beta+1}\left(1+\chi s^{-\zeta-\alpha}\right)}.
\end{split}
\end{equation}From now on, we only concentrate on the solution to $x(t)$.
Inverse Laplace transform of \eqref{eqs7} gives
\begin{equation*}\label{eqs10}\begin{split}
&x(t)=x_0 + v_0t^{\beta}E_{\alpha+\zeta,\beta+1}\left(-\chi t^{\alpha+\zeta}\right)\\&+\int_0^t (t-u)^{\alpha+\beta-1}E_{\alpha+\zeta, \alpha+\beta}\left(-\chi(t-u)^{\alpha+\zeta}\right)F(u)du,\end{split}
\end{equation*}
where
\begin{equation*}\label{eqs11}
E_{\mu,\nu}(z) =\sum_{k=0}^{\infty}\frac{z^k}{\Gamma(\mu k+\nu)}
\end{equation*}
is the two-parameter Mittag-Leffler function \cite{38}.  Taking expectation value using \eqref{eq33}, we have
\begin{equation*}\label{eqs13}
\left\langle x(t)\right\rangle = x_0 + v_0t^{\beta}E_{\alpha+\zeta,\beta+1}\left(-\chi t^{\alpha+\zeta}\right),
\end{equation*}
and
\begin{equation*}\label{eqs14}\begin{split}
\sigma_{xx}^2(t) &=\left\langle \left[x(t)-\langle x(t)\rangle\right]^2\right\rangle =2c_{\theta}\Gamma\left(1-\theta\right)\int_0^t u^{2\alpha+2\beta-\theta-1}\\&\times E_{\alpha+\zeta, \alpha+\beta}\left(-\chi u^{\alpha+\zeta}\right)E_{\alpha+\zeta,\alpha+\beta-\theta+1}\left( -\chi u^{\alpha+\zeta}\right)du,
\end{split}
\end{equation*}
where we have used the following identities \cite{27}: for  $\nu>0$ and $\theta\leq 1$,
\begin{equation*}\label{eqs15}\begin{split}&
\int_0^t (t-u)^{-\theta}u^{\nu-1}E_{\mu,\nu}\left(-\chi u^{\mu}\right)du\\ =& \Gamma(1-\theta) t^{\nu-\theta}E_{\mu, \nu-\theta+1}\left(-\chi t^{\mu}\right).\end{split}
\end{equation*}
The MSD is given by
\begin{equation*}\label{eqs17}\begin{split}
\bar{\Delta}^2(t) =& \sigma_{xx}^2(t)+\left(\left\langle x(t)\right\rangle-x_0\right)^2\\
=&\sigma_{xx}^2(t)+v_0^2t^{2\beta}E_{\alpha+\zeta,\beta+1}\left(-\chi t^{\alpha+\zeta}\right)^2.
\end{split}\end{equation*}

	Now by using the asymptotic properties of Mittag-Leffler function \cite{38}:
\begin{equation*}\label{eqs18}
\begin{split}
E_{\mu,\nu}(-z)=&\sum_{k=1}^N \frac{(-1)^{k-1}z^{-k}}{\Gamma\left(\nu-\mu k\right)}+O\left(z^{-1-N}\right), \hspace{0.5cm}z\rightarrow \infty,\\
E_{\mu,\nu}(-z)=&\frac{1}{\Gamma(\nu)}+O(z), \hspace{1cm}z\rightarrow 0,
\end{split}
\end{equation*}
one gets for $t\rightarrow 0$,
\begin{equation*}\label{eqs23}
\begin{split}
\sigma_{xx}^2(t) \sim & t^{2\alpha+2\beta-\theta}, \\
\bar{\Delta}^2(t)\sim &\begin{cases}
t^{2\beta}, \hspace{0.5cm}&\text{if}\;\; 2\alpha \geq \theta,\\
t^{2\alpha+2\beta-\theta}, \hspace{0.5cm}&\text{if}\;\; 2\alpha < \theta,
\end{cases}
\end{split}
\end{equation*}with the assumption that $2\alpha+2\beta>\theta$ so that $\sigma_{xx}^2(t)$ has a finite limit as $t\rightarrow 0$.

For large time asymptotic behaviors, we have generically
\begin{equation*}\label{eqs25}
\begin{split}
\sigma_{xx}^2(t) \sim &\begin{cases}
t^{2\beta-2\zeta-\theta}, \hspace{0.5cm}&\text{if}\;\; 2\beta-2\zeta>\theta,\\
\ln t, &\text{if}\;\; 2\beta-2\zeta=\theta,\\
\text{constant}, &\text{if}\;\; 2\beta-2\zeta<\theta.
\end{cases}\end{split}\end{equation*}Therefore, for the MSD, if $2\alpha\geq \theta$, the large time asymptotic of $\bar{\Delta}^2(t)$ is the same as $\sigma_{xx}^2(t) $. However, if $2\alpha<\theta$, then as $t\rightarrow \infty$,\begin{equation*}\begin{split}
\bar{\Delta}^2(t)\sim &\begin{cases}
t^{2\beta-2\zeta-2\alpha}, \hspace{0.5cm}&\text{if}\;\; 2\beta-2\zeta\geq 2\alpha,\\
\text{constant}, &\text{if}\;\; 2\beta-2\zeta<2\alpha.
\end{cases}
\end{split}
\end{equation*}
	 Now consider the general case with $0\leq \zeta< 1$. Fluctuation-dissipation theorem requires
\begin{equation*}\label{eqs26}
\begin{split}
\left\langle F(t)F(s) \right\rangle =c_{\theta}|t-s|^{-\theta}=k_BT\chi\frac{|t-s|^{\zeta-1}}{\Gamma(\zeta)},
\end{split}
\end{equation*}which gives $\theta=1-\zeta$. In order the MSD satisfies the properties of SFD, we require its asymptotic behavior $\sim t$ when $t\rightarrow 0$ and $\sim \sqrt{t}$ when $t\rightarrow \infty$. This gives two possibilities:
\begin{equation*}
\begin{split}
&\text{Case I} \;\; 2\alpha\geq 1-\zeta, \;\beta=\frac{1}{2}, \;2\beta-\zeta-1=\frac{1}{2},\hspace{2cm}\\
&\text{Case II}\;\; 2\alpha<1-\zeta,\;2\alpha+2\beta+\zeta-1=1,\\
&\hspace{2cm}2\beta-2\zeta-2\alpha=\frac{1}{2}.
\end{split}
\end{equation*}Case I implies that  $\zeta=-1/2$, which is a contradiction to $\zeta\in [0,1)$. For Case II, we find that \begin{equation}\label{eq4_16_3}
\alpha=\frac{3}{8}-\frac{3\zeta}{4}, \hspace{1cm}\beta =\frac{\zeta}{4}+\frac{5}{8}.\end{equation}The conditions $\alpha\in (0,1]$, $\beta\in (0, 1]$ and $2\alpha<1-\zeta$ imply that $\zeta\in [0, 1/2)$. In other words, for any $\zeta\in [0,1/2)$, define $\alpha$ and $\beta$ by \eqref{eq4_16_3}. Then the process $x(t)$ gives a correct description of SFD.

If we assume that $x(t)$ is ballistic instead of normal diffusive at small $t$, then the possibilities are
\begin{equation*}
\begin{split}
&\text{Case I} \;\; 2\alpha\geq 1-\zeta, \;\beta=1, \;2\beta-\zeta-1=\frac{1}{2},\hspace{2cm}\\
&\text{Case II}\;\; 2\alpha<1-\zeta,\;2\alpha+2\beta+\zeta-1=2,\\
&\hspace{2cm}2\beta-2\zeta-2\alpha=\frac{1}{2}.
\end{split}
\end{equation*}Case I gives $\beta=1$, $\zeta=1/2$ and $\alpha\geq 1/4$. For Case II, we find that
$$\alpha=\frac{5}{8}-\frac{3\zeta}{4}, \hspace{1cm}\beta =\frac{\zeta}{4}+\frac{7}{8}.$$ The condition $\beta\leq 1$ leads to $\zeta\leq 1/2$. However, the condition $2\alpha<1-\zeta$ gives $\zeta>1/2$, which is a contradiction. Therefore, in order that \eqref{eqs1} gives a suitable model for $x(t)$ which is ballistic at small $t$ and sub-diffusive of exponent $1/2$ at large $t$, we need to set $\beta=1$, $\zeta=1/2$ and $\alpha$ can be any number between $1/4$ and $1$. When $\alpha=1$, this reduces to the model \eqref{eq32} in the beginning of this section.

Finally, let us remark on the case where $v_0=0$. In this case, $\bar{\Delta}(t)^2=\sigma_{xx}^2(t)$. We then find that the properties of SFD are satisfied if $$\alpha=\frac{1}{4}-\zeta, \hspace{1cm}\beta=\frac{\zeta}{2}+\frac{3}{4}$$ for $\zeta\in [0, 1/4)$. When $\zeta=0$, this gives $\alpha=1/4$ and $\beta=3/4$, which agrees with the result in Section \ref{s3a}. On the other hand, if we require the process to be ballistic at small $t$, then $$\alpha=\frac{3}{4}-\zeta, \hspace{1cm}\beta=\frac{\zeta}{2}+\frac{3}{4}$$ for $\zeta\in [0, 3/4)$. $\zeta=0$ gives $\alpha=\beta=3/4$, which again agrees with the result in Section \ref{s3a}.
\subsection{ Extended Fractional Generalized  Langevin Equation}\label{s3c}
In this section,  we  stretch our Langevin approach to an even more general setting which includes the cases discussed in Section \ref{s3a} and Section \ref{s3b}. Consider the following extended version of the fractional  generalized Langevin equation:
\begin{equation}\label{eq4_16_4}\begin{split}
&_0D_t^{\beta}x(t)=v(t),\\
&_0D_t^{\alpha}v(t) +\lambda v(t) +\frac{\chi}{\Gamma(\zeta)}\int_0^t (t-u)^{\zeta-1}v(u)du=F(t),
\end{split}
\end{equation}
where now the dissipative memory  kernel $\gamma(t)$ is  given by
\begin{equation}\label{eq4_16_5}
\gamma(t) =2 \lambda \delta(t) + \frac{\chi}{\Gamma(\zeta)}t^{\zeta-1},
\end{equation}with $0\leq \zeta<1$. 	When $\alpha=\beta=1$ and $\zeta=1/2$, this system was proposed as a model to describe SFD for which the MSD is ballistic (i.e. $\sim t^2$) when $t$ is small.
When $\lambda=0$ or $\chi=0$, the equations \eqref{eq4_16_4} reduce to the equations \eqref{eqs1} considered in Section \ref{s3b}.  Therefore here we assume that $\lambda\neq 0$ and $\chi\neq 0$. As in Section III, taking Laplace transforms give
\begin{equation}\label{eq4_16_6}\begin{split}
\tilde{x}(s) =&\frac{x_0}{s}+\frac{\tilde{F}(s)}{s^{\alpha+\beta}\left(1+\lambda s^{-\alpha}+\chi s^{-\zeta-\alpha}\right)}\\&+\frac{v_0}{s^{\beta+1}\left(1+\lambda s^{-\alpha}+\chi s^{-\zeta-\alpha}\right)}.
\end{split}\end{equation} In the following, we assume that $v_0\neq 0$. Let $K_1(t)$ and $K_2(t)$ be respectively the inverse Laplace transforms of
$$	\frac{1}{s^{\alpha+\beta}\left(1+\lambda s^{-\alpha}+\chi s^{-\zeta-\alpha}\right)}\;\text{and}\;\frac{1}{s^{\beta+1}\left(1+\lambda s^{-\alpha}+\chi s^{-\zeta-\alpha}\right)}.$$ Then the inverse Laplace transform of \eqref{eq4_16_6} gives
\begin{equation}\label{eq4_16_7}
x(t) = x_0 +\int_0^t K_1(t-u)F(u)du + v_0K_2(t).
\end{equation}From this, we find that the variance of $x(t)$ and the  MSD is given respectively by
\begin{equation}\label{eq4_16_8}
\begin{split}
\sigma_{xx}^2(t) =&2\int_0^t \int_0^u K_1(v) C_F(u-v) K_1(u)dvdu\\\bar{\Delta}^2(t)=&2\int_0^t \int_0^u K_1(v) C_F(u-v) K_1(u)dvdu+ v_0^2K_2(t)^2.
\end{split}
\end{equation}
The asymptotic behaviors of the functions 	$K_1(t)$ and $K_2(t)$ at small and large $t$ are studied in Appendix \ref{a1}. The result is: as $t\rightarrow 0$,
\begin{equation}\label{eq4_16_9}
K_1(t) \sim t^{\alpha+\beta-1}, \hspace{1cm} K_2(t)\sim t^{\beta}.
\end{equation}As $t\rightarrow \infty$,
\begin{equation}\label{eq4_16_10}
\begin{split}
K_1(t)\sim t^{\beta-\zeta-1},\hspace{1cm} K_2(t)\sim t^{\beta-\zeta-\alpha}.
\end{split}
\end{equation}As in Section \ref{s3b}, we assume the following generalized fluctuation-dissipation theorem \cite{4_17_1}:
\begin{equation}\label{eq4_16_11}
C_F(t)=k_BT\left(2 \lambda \delta(t) + \frac{\chi}{\Gamma(\zeta)}t^{\zeta-1}\right).
\end{equation}Then the MSD \eqref{eq4_16_8} can be rewritten as
\begin{equation}\label{eq4_16_12}
\begin{split}&\bar{\Delta}^2(t) = v_0^2K_2(t)^2+2k_BT\lambda \int_0^t  K_1(u)^2du\\&+ 2k_BT\frac{\chi}{\Gamma(
\zeta)}\int_0^t \int_0^u K_1(v) (u-v)^{\zeta-1}K_1(u) dvdu.
\end{split}
\end{equation}
\eqref{eq4_16_9} then implies that its small time asymptotic behavior is
\begin{equation*}\label{eq4_16_13}
\bar{\Delta}^2(t)\sim \begin{cases}t^{ 2\beta}, \hspace{0.5cm}&\text{if}\;\; \alpha\geq 1/2,\\
t^{2\alpha+2\beta-1}, &\text{if}\;\;\alpha<1/2.
\end{cases}
\end{equation*}For the large time asymptotic behavior of the MSD, \eqref{eq4_16_10} and \eqref{eq4_16_12} give
\begin{equation}\label{eq4_16_14}
\begin{split}
&\bar{\Delta}^2(t) \sim   \begin{cases}t^{2\beta-\zeta-1}, \hspace{0.5cm}&\text{if}\;\;2\alpha+\zeta\geq 1\; \text{and}\; 2\beta-\zeta>1\\\ln t, &\text{if}\;\;2\alpha+\zeta\geq 1\; \text{and}\; 2\beta-\zeta=1\\
\text{constant},&\text{if}\;\;2\alpha+\zeta\geq 1\; \text{and}\; 2\beta-\zeta<1\\
t^{2\beta-2\zeta-2\alpha}, &\text{if}\;\; 2\alpha+\zeta<1\;\text{and}\; \beta-\zeta\geq\alpha \\
\text{constant}, &\text{if}\;\; 2\alpha+\zeta<1\;\text{and}\; \beta-\zeta<\alpha\end{cases}.
\end{split}
\end{equation}The short time asymptotic behavior of the MSD is governed by the term $\delta(t)$ in the dissipative memory  kernel and the long time asymptotic behavior is governed by the term $t^{\zeta-1}$. This gives the general form of dissipative memory kernel \eqref{eq4_16_5} the advantage of being able to interpolate between the two particular cases with $\lambda=0$ and $\chi=0$ considered in the previous subsections. Now to satisfy the characteristics of SFD, there are a few possibilities:
\begin{equation*}
\begin{split}
&\text{Case I} \;\; \alpha\geq \frac{1}{2}, \;\beta =\frac{1}{2},\;2\beta-\zeta-1=\frac{1}{2},\\
&\text{Case II}\;\; \alpha<\frac{1}{2},\; 2\alpha+2\beta-1=1, \;2\alpha+\zeta\geq 1,\\&\hspace{2cm}\; 2\beta-\zeta-1=\frac{1}{2},\hspace{8cm}\\
&\text{Case III}\;\; \alpha<\frac{1}{2},\; 2\alpha+2\beta-1=1, \;2\alpha+\zeta <1,\\&\hspace{2cm}\; 2\beta-2\zeta-2\alpha=\frac{1}{2}.
\end{split}
\end{equation*}We have used the fact that $\alpha\geq 1/2$ implies $2\alpha\geq 1>\zeta$. Case I gives     $\zeta=-1/2$, which is a contradiction. The two equalities in Case II imply that $2\alpha+\zeta=1/2$, which contradicts $2\alpha+\zeta\geq 1$. For Case III, the two equalities lead to
\begin{equation}\label{eq4_16_15}
\alpha=\frac{3}{8}-\frac{\zeta}{2},\hspace{1cm}\beta=\frac{5}{8}+\frac{\zeta}{2}.
\end{equation}The condition $2\alpha+\zeta<1$ is then automatically satisfied. The conditions $0<\alpha<1/2$ and $0<\beta\leq 1$ then imply that $0\leq \zeta<3/4$. In other words, for any $\zeta\in [0, 3/4)$, define $\alpha$ and $\beta$ by \eqref{eq4_16_15}. Then we obtain a solution $x(t)$ to \eqref{eq4_16_4} which has the characteristics of SFD. Compared to the solution in Section \ref{s3b} which only allow $\zeta$ to lie in the range $[0,1/2)$, we find that the extended  fractional generalized Langevin equation \eqref{eq4_16_4} can be used to describe SFD in a larger range of $\zeta$. In both cases,  the maximum value of $\alpha$ is $3/8$.

Next we consider the case addressed in the paper \cite{4_17_1}, where the MSD is ballistic for small $t$, and become sub-diffusive $\sim \sqrt{t}$ when $t$ is large enough. For this to happen, there are a few possibilities:
\begin{equation*}
\begin{split}
&\text{Case I} \;\; \alpha\geq \frac{1}{2}, \;\beta =1,\;2\beta-\zeta-1=\frac{1}{2},\\
&\text{Case II}\;\; \alpha<\frac{1}{2},\; 2\alpha+2\beta-1=2, \;2\alpha+\zeta\geq 1,\\&\hspace{2cm}\; 2\beta-\zeta-1=\frac{1}{2},\hspace{8cm}\\
&\text{Case III}\;\; \alpha<\frac{1}{2},\; 2\alpha+2\beta-1=2, \;2\alpha+\zeta <1,\\&\hspace{2cm}\; 2\beta-2\zeta-2\alpha=\frac{1}{2}.
\end{split}
\end{equation*}For the first case, we find that $\beta=1$, $\zeta=1/2$ and the only restriction on $\alpha$ is $\alpha\geq 1/2$. When $\alpha=1$, this is the case considered in \cite{4_17_1}. Case II and Case III imply that $\alpha=3/2-\beta\geq 1/2$ since $\beta\leq 1$. But this violates the condition $\alpha<1/2$. In summary, to characterize the behavior of a system which is ballistic at small $t$ and sub-diffusive at large $t$, one can use the extended generalized fractional Langevin model \eqref{eq4_16_4} with $\beta=1$ (so that the velocity is normal derivative of position), $\zeta=1/2$ and $\alpha$ any values between $1/2$ and $1$. Compare to the result of previous section, we find that when $\lambda=0$, the conditions $\zeta=1/2$ and $\beta=1$ are the same, but the range of $\alpha $ is from $1/4$ to $1$, which has larger range of values as compared to the case when $\lambda\neq 0$.

Finally we remark on the case where $v_0=0$. In this case, $\bar{\Delta}^2(t)=\sigma_{xx}^2(t)$. Therefore, the small and large time asymptotic behaviors of the MSD for SFD is satisfied if $$\alpha=\frac{1}{4}-\frac{\zeta}{2},\hspace{0.5cm} \beta=\frac{\zeta}{2}+\frac{3}{4}, \hspace{0.5cm}0\leq \zeta <\frac{1}{2};$$ and the MSD is ballistic at small $t$ if
$$\alpha=\frac{3}{4}-\frac{\zeta}{2},\hspace{0.5cm} \beta=\frac{\zeta}{2}+\frac{3}{4}, \hspace{0.5cm}0\leq \zeta <1.$$

\subsection{Further Generalizations}
\begin{figure}\centering \epsfxsize=0.49\linewidth
 \epsffile{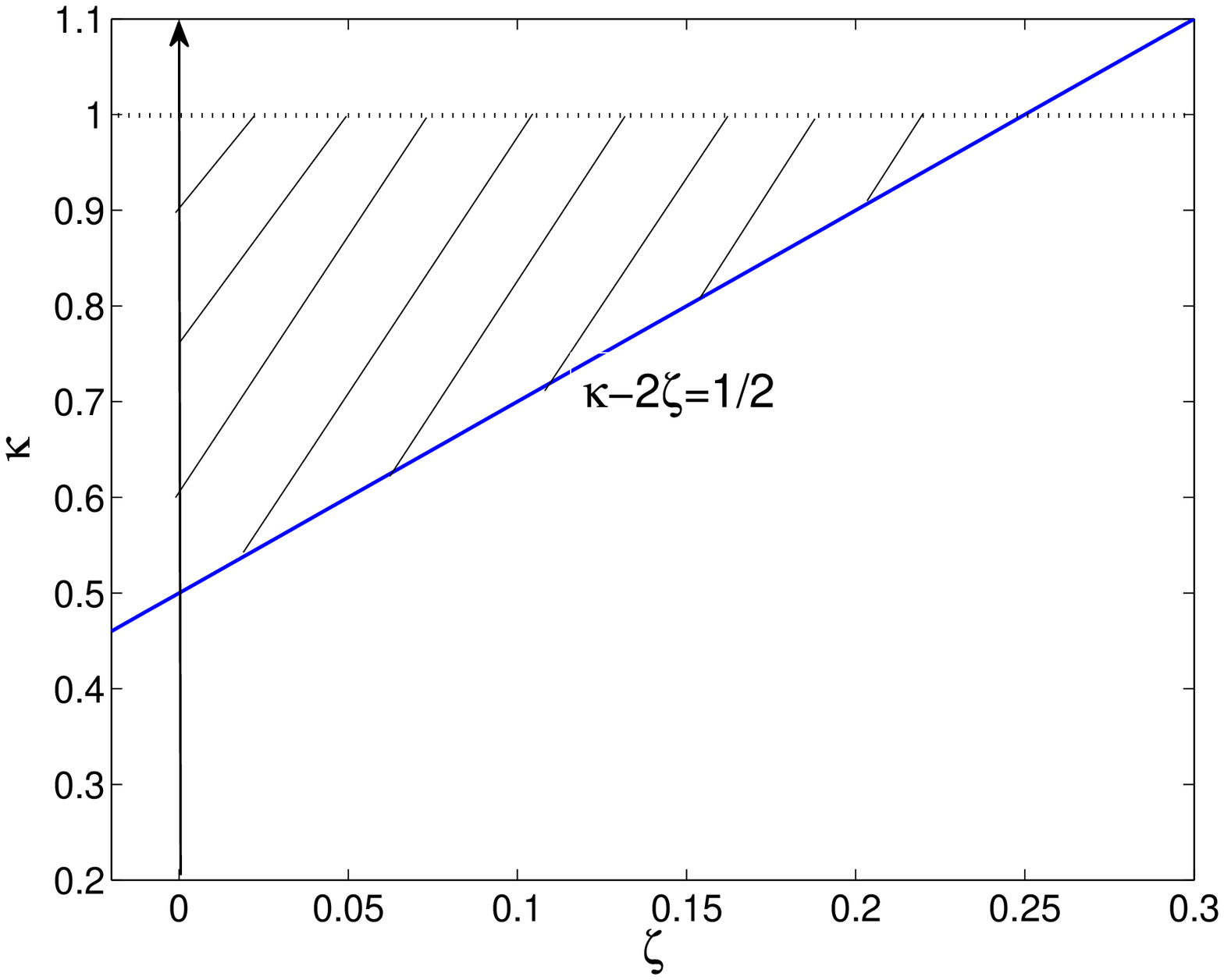} \centering \epsfxsize=0.49\linewidth
 \epsffile{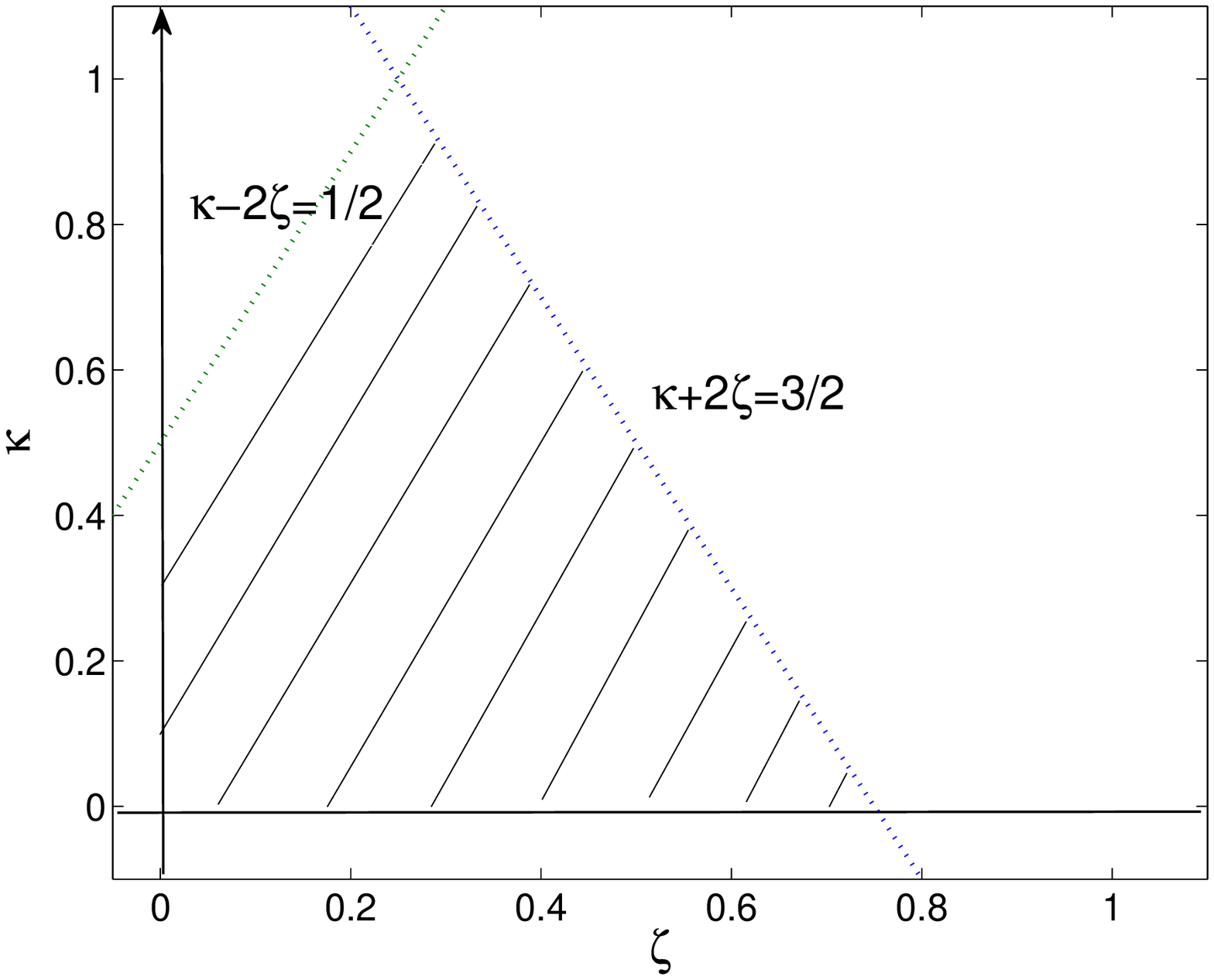}\caption{\label{f2}  The feasible regions for $(\zeta,\kappa)$ if the MSD is diffusive at small time. }  \end{figure}

 The results above can be easily generalized to fractional Langevin equation with  more general (nonlocal) dissipative memory  kernel $\gamma(t)$ which behaves like $\sim t^{\kappa-1}$, $\kappa \in [0,1)$ when $t\rightarrow 0$ and behaves like $\sim t^{\zeta-1}$, $\zeta\in [0,1)$ when $t\rightarrow \infty$. We note that $\zeta=\kappa=0$ is the special case of the Dirac delta function kernel. The  fractional generalized Langevin equation considered in Section \ref{s3b} corresponds to $\kappa=\zeta$, whereas the extended  fractional generalized Langevin equation considered in Section \ref{s3c} corresponds to $\kappa=0$. For the Laplace transform of the dissipative memory  kernel $\tilde{\gamma}(s)$, one finds that $\tilde{\gamma}(s)\sim s^{-\zeta}$ when $s\rightarrow 0$ and $\tilde{\gamma}(s)\sim s^{-\kappa}$ when $s\rightarrow \infty$. The solution $x(t)$ can be written as \eqref{eq4_16_7}, where the small-$t$ and large-$t$ asymptotic behaviors of the functions $K_1(t)$ and $K_2(t)$ are still given by \eqref{eq4_16_9} and \eqref{eq4_16_10}. The generalized fluctuation-dissipation theorem \eqref{eq4_16_11} then implies that similar to $\gamma(t)$,  $C_F(t)\sim t^{\kappa-1}$ when $t\rightarrow 0$ and $C_F(t)\sim t^{\zeta-1}$, when $t\rightarrow \infty$. One can then deduce that when $t\rightarrow 0$, the MSD \eqref{eq4_16_8} behaves like \begin{equation*}
\bar{\Delta}^2(t)\sim \begin{cases}t^{ 2\beta}, \hspace{0.5cm}&\text{if}\;\; 2\alpha +\kappa \geq 1,\\
t^{2\alpha+2\beta+\kappa-1}, &\text{if}\;\;2\alpha+\kappa <1.
\end{cases}
\end{equation*}The large-time asymptotic of the MSD is still given  by \eqref{eq4_16_14}. After some analysis, we find that the short and long time properties of SFD are satisfied if $(\zeta, \kappa)$ satisfies (see the first graph in FIG. \ref{f2}) $$\zeta\geq 0, \;\; \kappa-2\zeta\geq \frac{1}{2}, \;\;\kappa<1,$$ and the values of $\alpha$ and $\beta$ are $$\alpha=\frac{1}{4}-\zeta, \;\;\beta=\frac{1}{2};$$or
$(\zeta, \kappa)$ satisfies (see the second graph in FIG. \ref{f2}) \begin{equation}\label{eq4_17_5}\zeta\geq 0, \;\; \kappa\geq 0,\;\;\kappa-2\zeta< \frac{1}{2}, \;\;\kappa+2\zeta<\frac{3}{2},\end{equation} and the values of $\alpha$ and $\beta$ are $$\alpha=\frac{3}{8}-\frac{\kappa}{4}-\frac{\zeta}{2}, \;\;\beta=\frac{5}{8}-\frac{\kappa}{4}+\frac{\zeta}{2}.$$

\begin{figure}\centering \epsfxsize=0.6\linewidth
 \epsffile{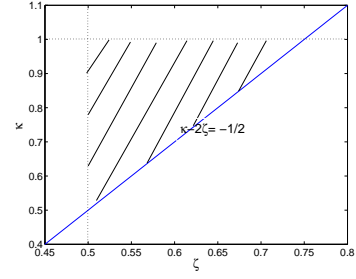}  \caption{\label{f3}  The feasible region for $(\zeta,\kappa)$ if the MSD is ballistic at small time. }  \end{figure}

On the other hand,  the MSD is $\sim t^2$ (ballistic) at small $t$ if $\kappa \in [0, 1)$, $\beta=1$, $\zeta=1/2$ and $$\alpha \geq \max\left\{\frac{1}{4}, \frac{1-\kappa}{2}\right\};$$ or $(\zeta,\kappa)$ satisfies (see FIG. \ref{f3})
\begin{equation}\label{eq4_17_6}\frac{1}{2}<\zeta<\frac{3}{4}, \;\;\kappa-2\zeta\geq -1/2,\;\;\kappa<1,\end{equation}and
$$\alpha=\frac{3}{4}-\zeta,\;\;\beta=1.$$

Finally we would like to comment that if the dissipative memory kernel is a finite sum of power-law functions, then $\kappa\leq \zeta$, and the situations \eqref{eq4_17_5} and \eqref{eq4_17_6} would not appear.
\section{ Concluding Remarks}
We have proposed some stochastic processes that can   describe the basic properties of SFD. In the first case, we find that the replacement of the Hurst exponent in fractional Brownian motion by a two-valued step function can be used to model SFD that is normal diffusive at small time and sub-diffusive at large time. Since fractional Brownian motion can be considered as a   solution of a special case of fractional Langevin equation, it is natural to consider modeling SFD by fractional Langevin equation. We have discussed in detail the cases where the dissipative memory  kernel is a Dirac delta function, a   power-law function as well as the combination of these two. For each of these cases, we find that there is a range of the parameters where the corresponding fractional Langevin equation can be used to model process whose MSD behaves like $\sim t$ when $t$ is small and behaves like $\sim \sqrt{t}$ when $t$ is large. This range has been  determined explicitly. The corresponding range of the parameters for which the MSD of the process behaves like $\sim t^2$ instead of $\sim t$ when $t$ is small  is also determined.

  One weakness of usig SFBM to model SFD is that the change of scaling exponent is not continuous so that the transition from the normal diffusion to the anomalous subdiffusion occurs abruptly. This weakness is overcome by various fractional Langevin models. However, we only use the short and long time behaviors of the stochastic processes to determine  the model.  For a more complete description of SFD, it is necessary to take into account the boundary conditions at the ends of the single-file system, as these would become relevant if the system is of finite extension. Furthermore, one may also have to consider the interaction between the diffusion particles with the wall of the single-file system. Thus, the evolution of the particles in the single-file system during the intermediate time interval becomes relevant. We hope to incorporate the physical mechanism of SFD in order to obtain a more realistic model of SFD in our future work.

\begin{acknowledgments}
The authors would like to thank the Malaysian Ministry of Science, Innovation and Technology for funding this project under eScienceFund.
\end{acknowledgments}

\appendix
\section{The functions $\boldsymbol{K}_{\boldsymbol{1}}\boldsymbol{(t)}$, $\boldsymbol{K}_{\boldsymbol{2}}\boldsymbol{(t)}$ and their asymptotic properties}\label{a1}
In this appendix, we give the details of the analysis of the asymptotic properties of the functions $K_1(t)$ and $K_2(t)$ defined as the inverse Laplace transform of
$$	\frac{1}{s^{\alpha+\beta}\left(1+\lambda s^{-\alpha}+\chi s^{-\zeta-\alpha}\right)}\;\;\text{and}\;\;\frac{1}{s^{\beta+1}\left(1+\lambda s^{-\alpha}+\chi s^{-\zeta-\alpha}\right)}.$$By taking Laplace transforms, it is easy to check that
\begin{equation*}\label{eqa1}\begin{split}
K_1(t)=&\sum_{k=0}^{\infty}(-1)^k \sum_{j=0}^k \begin{pmatrix} k\\j\end{pmatrix} \lambda^{k-j}\chi^j \frac{t^{\alpha k +\zeta j +\alpha+\beta-1}}{\Gamma\left(\alpha k +\zeta j +\alpha+\beta\right)},\\
K_2(t)=&\sum_{k=0}^{\infty}(-1)^k \sum_{j=0}^k \begin{pmatrix} k\\j\end{pmatrix} \lambda^{k-j}\chi^j \frac{t^{\alpha k +\zeta j  +\beta }}{\Gamma\left(\alpha k +\zeta j +\beta+1\right)}.\end{split}\end{equation*}Therefore, as $t\rightarrow 0$,
\begin{equation*}\label{eqa2}
K_1(t)\sim t^{\alpha+\beta-1}\;\;\text{and}\;\; K_2(t)\sim t^{\beta}.
\end{equation*}This behavior is independent of the coefficients $\lambda$ and $\chi$. For the large-$t$ asymptotic behaviors, notice that
\begin{equation}\label{eqa3}\begin{split}
K_1(t)=&\frac{1}{\Gamma(\alpha+\beta)}\int_0^t (t-u)^{\alpha+\beta-1}K_0(u)du,\\
K_2(t)=&\frac{1}{\Gamma( \beta+1)}\int_0^t (t-u)^{ \beta }K_0(u)du,
\end{split}
\end{equation}where $K_0(t)$ is the inverse Laplace transform of
$$	\frac{1}{ 1+\lambda s^{-\alpha}+\chi s^{-\zeta-\alpha} },$$ which has an integral representation
\begin{equation*}\label{eqa4}
\begin{split}
K_0(t)=&\frac{1}{2\pi i}\int_{-i\infty}^{i\infty}\frac{e^{st}}{ 1+\lambda s^{-\alpha}+\chi s^{-\zeta-\alpha} }ds\\
=&\frac{1}{2\pi }\int_{0}^{\infty}\frac{e^{ist}}{ 1+\lambda (is)^{-\alpha}+\chi (is)^{-\zeta-\alpha} }ds\\
&+\frac{1}{2\pi }\int_{0}^{\infty}\frac{e^{-ist}}{ 1+\lambda (-is)^{-\alpha}+\chi (-is)^{-\zeta-\alpha} }ds\\
=&\frac{\text{Im}}{\pi}\int_0^{\infty}\frac{e^{-st}}{1+ \lambda e^{i\pi\alpha}s^{-\alpha}+\chi e^{i\pi(\zeta+\alpha)}s^{-\zeta-\alpha}}ds\\
=&\frac{\text{Im}}{\pi}\int_0^{\infty}\frac{s^{\zeta+\alpha}e^{-st}}{s^{\zeta+\alpha}+ \lambda e^{i\pi\alpha}s^{\zeta}+\chi e^{i\pi(\zeta+\alpha)} }ds.
\end{split}
\end{equation*}Now for the large-$t$ asymptotic behavior of $K_0(t)$, we have
\begin{equation*}\label{eqa5}
\begin{split}
&K_0(t) =t^{-1-\zeta-\alpha}\frac{\text{Im}}{\pi}\int_0^{\infty}\frac{s^{\zeta+\alpha}e^{-s}}{\left(\frac{s}{t}\right)^{\zeta+\alpha}+ \lambda e^{i\pi\alpha}\left(\frac{s}{t}\right)^{\zeta}+\chi e^{i\pi(\zeta+\alpha)} }ds\\
\end{split}\end{equation*}\begin{equation*}\begin{split}=&t^{-1-\zeta-\alpha}\frac{\text{Im}}{\pi}\int_0^{\infty}\sum_{k=0}^{\infty}\Biggl\{\frac{(-1)^k}{\chi^{k+1}}\sum_{j=0}^k \begin{pmatrix}k\\j\end{pmatrix}\lambda^{k-j} e^{-i\pi(\zeta k +\alpha j+\zeta+\alpha)}\Biggr\}\\
&\hspace{2cm}\times t^{-\zeta k-\alpha j} s^{\zeta k+\alpha j +\zeta+\alpha}e^{-s}ds\\
\sim & -\frac{t^{-1-\zeta-\alpha}}{\pi}\sum_{k=0}^{\infty}\Biggl\{\frac{(-1)^k}{\chi^{k+1}}\sum_{j=0}^k \begin{pmatrix}k\\j\end{pmatrix}\lambda^{k-j} \sin \left(\pi(\zeta k +\alpha j+\zeta+\alpha)\right)\\&\hspace{2cm}\times \Gamma(\zeta k+\alpha j +\zeta+\alpha+1) t^{-\zeta k-\alpha j}\Biggr\}.
\end{split}
\end{equation*}This implies that generically,
\begin{equation*}\label{eqa6}
K_0(t)\sim t^{-1-\zeta-\alpha},\;\;\;\;t\rightarrow \infty.
\end{equation*}We then obtain from \eqref{eqa3} that as $t\rightarrow \infty$,
\begin{equation*}\label{eqa7}
\begin{split}
K_1(t)\sim t^{\beta-\zeta-1},\;\;\; K_2(t)\sim t^{\beta-\zeta-\alpha}.
\end{split}
\end{equation*}

\end{document}